\RequirePackage{fix-cm}
\documentclass[smallextended]{svjour3}       % onecolumn (second format)
\smartqed  % flush right qed marks, e.g. at end of proof
\usepackage{graphicx}
\usepackage{amsmath}
\usepackage{algorithm}

\usepackage{subcaption}
\usepackage[noend]{algpseudocode}
\usepackage{ifluatex}
\usepackage{tikz-qtree}
\usepackage{xy}
\tikzstyle{vertex}=[circle, draw, inner sep=0pt, minimum size=15pt]
\usetikzlibrary{backgrounds,fit,decorations.pathreplacing}  % TikZ libraries
\usetikzlibrary{shapes,arrows}
\input{Qcircuit}
\newcommand{\ket}[1]{\ensuremath{\left|#1\right\rangle}} % Dirac Kets
\newcommand{\vertexs}{\node[vertexs]}
\newcommand{\vertex}{\node[vertex]}
\begin{document}

\title{A dynamic programming approach for distributing quantum circuits by bipartite graphs
}

\author{Zohreh Davarzani\and Mariam Zomorodi-Moghadam \and Mahboobeh Houshmand  \and Mostafa Nouri-baygi}

%\authorrunning{Short form of author list} % if too long for running head

\institute{Z. Davarzani \at
            Department of Computer Engineering, Ferdowsi University of Mashhad, Mashhad, Iran \\
                     %  \\
%             \emph{Present address:} of F. Author  %  if needed
           \and
           M. Zomorodi-Moghadam \at 
              Department of Computer Engineering, Ferdowsi University of Mashhad, Mashhad, Iran \\
              \email{m$\_$zomorodi@um.ac.ir}           %  \\
%             \emph{Present address:} of F. Author  %  if needed
   \and
          M. Houshmand \at
                  Department of Computer Engineering, Mashhad Branch, Islamic Azad University, Mashhad, Iran\\
                     %  \\
%             \emph{Present address:} of F. Author  %  if needed
         \and   
           M. Nouri-baygi \at
               Department of Computer Engineering, Ferdowsi University of Mashhad, Mashhad, Iran\\
                       %  \\
}

\date{Received: date / Accepted: date}
% The correct dates will be entered by the editor
\maketitle

\begin{abstract}
Near-term large quantum computers are not able to operate as a single processing unit. It is therefore required to partition a quantum circuit into smaller parts, and then each part is executed on a small unit. This approach is known as distributed quantum computation. In this study, a dynamic programming algorithm is proposed to minimize the number of communications in a distributed quantum circuit (DQC). This algorithm consists of two steps: first, the quantum circuit is converted into a bipartite graph model, and then a dynamic programming approach (DP) is proposed to partition the model into low-capacity quantum circuits. The proposed approach is evaluated on some benchmark quantum circuits with remarkable reduction in the number of required teleportations.
% \PACS{PACS code1 \and PACS code2 \and more}
% \subclass{MSC code1 \and MSC code2 \and more}
\end{abstract}

\section{Introduction}
\label{intro}
Nowadays, with the empirical demonstrations of quantum computing, this field has witnessed a rapid growth with high performance in many areas such as database searching and integer factorization and etc. Quantum computation has many advantages over classical ones, but having a large-scale quantum system with many qubits, has implementation constraints \cite{Zomorodi} which makes distributed quantum implementation a necessity \cite{van2010}. One challenge in distributed quantum computation is the interconnection between qubits and the environment, which makes quantum information more delicate and leads to error \cite{krojanski2004scaling}. Distributed quantum system overcomes these problems in the sense that qubits are distributed to subsystems and each one is responsible of computation between fewer qubits. Therefore instead of having a large-scale quantum computer, it is reasonable and better to have a set of the limited-capacity quantum system which interact within quantum or classical channel and built the behavior of whole quantum system\cite{nickerson2013topological}. This concept is known as distributed quantum system.

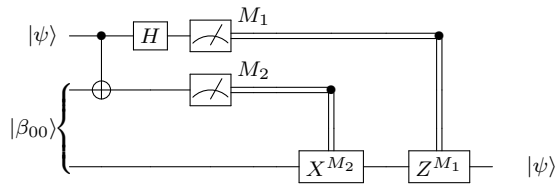
\begin{figure}[ht]

\centering
\[
\Qcircuit @C=1em @R=1em{
\lstick{\ket{\psi}} & \ctrl{1} & \gate{H} &\meter &\ustick{M_1} \cw  & \cw & \cw & \cw & \control \cw & & \\
 & \targ & \qw & \meter & \ustick{M_2} \cw & \cw &  \control \cw & &\cwx & & \\
\lstick{\ket{\beta_{00}}} & & & & & & \cwx & &\cwx & & \\
 & \qw &\qw & \qw & \qw & \qw & \gate{X^{M_2}} \cwx & \qw & \gate{Z^{M_1}} \cwx & \qw & \rstick{\ket{\psi}} \gategroup{2}{1}{4}{1}{.5em}{\{}
}
\]
\caption{\small{Quantum circuit for teleporting a qubit~\cite{Nielsen10}.}}
\label{fig1}
\end{figure}

DQC architecture can be described as follows \cite {Pablo}:
\begin{itemize}
\item
\textbf{} Multiple quantum processing units (QPUs), each unit keeps a number of qubits and can execute some universal quantum gates on them.
\item
\textbf{}A classical communication network, the QPUs send or receive messages through this network when measuring their qubits. 
\item
\textbf{}Ebit generation hardware, ebit is shared between two QPUs and consists of two qubits. Each of them is placed in different QPU.
Also, an ebit includes the required information for sending a single qubit from one QPU to another one. Each QPU may have the hardware to generate and share ebits or they may be created by a central device.
\end{itemize}
A Distributed Quantum Circuit (DQC) consists of $K$ smaller quantum circuits (called partition) with fewer qubits and limited capacity where partitions are far from each other \cite{Meter06,meter}. It is necessary for DQC to have a reliable protocol for interconnection between subsystems.
Teleportation \cite {Bennett} is a primitive protocol for interconnection between qubits by using entanglement of qubits, which is led to distribution of information through quantum system \cite {Whit}. Figure \ref{fig1} shows the quantum circuit for basic teleportation, as described in \cite {Nielsen10}. In this figure, two top lines are the sender's qubits and the bottom line is the receiver’s one.
In this protocol, qubits transfer their states from one point to another without moving them physically. Finally, they perform computations locally on qubits. This approach is called teledata. There is another approach which is called telegate. In \cite{van2010}, telegate and teledata are discussed. In telegate, gates are executed remotely using the teleported gate without needing qubits to be nearby. Authors have shown that teledata is more appropriate for DQC systems and have used teledata for building a DQC system out of a monolithic quantum circuit.
Teleportation is an expensive operation in DQC. Also according to no-cloning theorem \cite{wootteres}, when a qubit teleports its state to a destination, after a while it may be required in its subsystem again. Therefore, it is essential to minimize the number of teleportations in DQC.
Dynamic programming (DP) is one important method for mathematical optimization and computer sciences and is widely used in many fields. In DP approach, the main problem is decomposed into smaller sub-problems and once all the sub-problems have been solved, one optimal solution to the large problem is left.
In this paper, an algorithm is proposed to solve the problem of quantum circuit distribution. The algorithm consists of two steps: in the first step quantum circuit is modeled with a bipartite graph and in the next step, a dynamic programming approach is presented to partition the bipartite graph into $K$ parts in the sense that the number of connections between the parts is minimized.

The paper is organized as follows. In Section 2, some definitions and notations of distributed Quantum computing are described. Related works are presented in Section 3. In Section 4, partitioning of the bigraph is described. The proposed algorithm is presented in Section 5 and finally experimental results for some benchmarks are presented in Section 6.

\section{Definitions and notations}

In quantum computing, a qubit is the basic unit of quantum information. A qubit is a two-level quantum system and its state can be represented by a unit vector in a two-dimensional Hilbert space for which an orthogonal basis set denoted by $\{\ket{0},\ket{1}\}$ has been fixed. Qubits can be in a superposition of $\ket{0}$ and $\ket{1}$ in form of $\alpha\ket{0}+\beta\ket{1}$ where $\alpha$ and $\beta$ are complex numbers such that $\mid\alpha\mid^2+\mid\beta\mid^2=1$. When the qubit state is measured, with probabilities $\mid\alpha\mid^2$ and $\mid\beta\mid^2$, classical outcomes of 0 and 1 are observed respectively.

There are many ways to present a quantum algorithm. For example adiabatic model of computation~\cite{farhi2000quantum} and quantum programming languages~\cite{zomorodi2014synthesis}. But one of the mostly used approaches is quantum circuit \cite{deutsch}: a model for quantum computation by a sequence of quantum gates to transfer information on the input quantum registers. The quantum circuit is based on unitary evolution by networks of these gates \cite{Nielsen10}. Every $n$-qubit quantum gate is a linear transformation represented by a unitary matrix on an $n$-qubit Hilbert space. A set of useful single-qubit gates called Pauli set are defined bellow \cite{weinstein2014steane}, \cite{zomorodi2016rotation}:

\begin{equation}
\sigma_{0}=I=
\begin{bmatrix}
1&0\\
0&1\\
\end{bmatrix}
\end{equation}
\begin{equation}
\sigma_{1}=X=
\begin{bmatrix}
0&1\\
1&0\\
\end{bmatrix}
\end{equation}
\begin{equation}
\sigma_{2}=Y=
\begin{bmatrix}
0&-i\\
i&0\\
\end{bmatrix}
\end{equation}
\begin{equation}
\sigma_{3}=Z=
\begin{bmatrix}
1&0\\
0&-1\\
\end{bmatrix}
\end{equation}
Another important single qubit gate is Hadamard which is defined as:
\begin{equation}
H=1/\sqrt{2}
\begin{bmatrix}
1&1\\
1&-1\\
\end{bmatrix}
\end{equation}

\begin{figure}
\begin{center}
\includegraphics[scale=0.3]{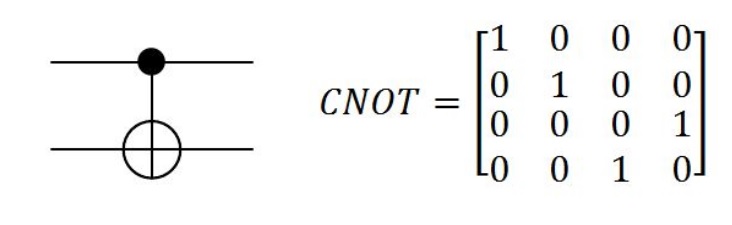}
\caption{\small{Circuit and Matrix representation of CNOT gate}}
\label{fig2}
\end{center}
\end{figure}

A controlled-$U$ is a two-qubit gate which acts on two qubits, namely, control and target qubits. When the control qubit is $\ket{1}$, $U$ is applied to the target qubit, otherwise the target qubit remains unchanged. One of the most useful controlled-$U$ gates is controlled-Not (CNOT) gate. This gate applies operator $X$ to the target qubit, if the control qubit is $\ket{1}$. Otherwise the target qubit do not change.
Figure \ref{fig2} shows the circuit and the matrix representation of the CNOT gate.

\begin{figure}[h]
\begin{center}
    \begin{tikzpicture}
    %
    % `operator' will only be used by Hadamard (H) gates here.
    % `phase' is used for controlled phase gates (dots).
    % `surround' is used for the background box.
    \tikzstyle{operator} = [draw,shape=circle,fill=white,minimum size=1.5em] 
    \tikzstyle{phase} = [draw,shape=circle,minimum size=10pt,inner sep=0pt]
    \tikzstyle{control} = [fill,shape=circle,minimum size=7pt,inner sep=0pt]

    %
    % Qubits
    \node at (0,0) (q1) {\ket{q_{1}}};
    \node at (0,-1) (q2) {\ket{q_{2}}};
    \node at (0,-2) (q3) {\ket{q_{3}}};

    %
    % Column 2
    \node[control] (c20) at (2,0) {} ;%edge [-] (q1);
    \node[operator] (r22) at (2,-2) {} ;%edge [-] (q3);
    \draw[-] (2,-2.25) -- (c20);

 % Column 3
    \node[control] (c30) at (3,0) {}; %edge [-] (c20);
    \node[operator] (r32) at (3,-2) {}; %edge [-] (r22);
   \draw[-] (c30) -- (3,-2.25);
 % Column 4

    \node[control] (r42) at (4,-2) {} ;%edge [-] (r32);
    \node[operator] (c41) at (4,-1) {};% edge [-] (q2);
    \draw[-] (4,-0.75) -- (r42);

 \node (end1) at (5,0) {}; %edge [-] (c30);
\node (end2) at (5,-1) {} ;%edge [-] (c41);
 \node (end3) at (5,-2) {} edge [-] (r42);

 \node at (5.5,0) (q1) {\ket{q_{1}}};
    \node at (6.5,-1) (q2) {$\ket{q_{1}}\bigoplus \ket{q_{2}} \bigoplus \ket{q_{3}}$};
    \node at (6,-2) (q3) {\ket{q_{3}} $\bigoplus $ \ket{q_{2}}};
\draw[-](1,0)--(5,0);
\draw[-](1,-1)--(5,-1);
\draw[-](1,-2)--(5,-2);
    \end{tikzpicture}
\caption{\small{A sample quantum circuit }}
\label{fig3}
\end{center}
\end{figure}
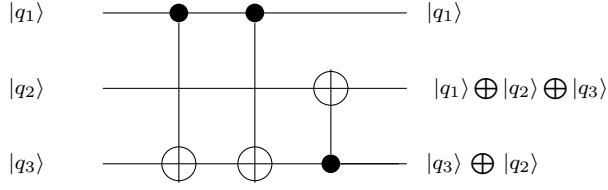

A quantum circuit consists of several quantum gates interacted by quantum wires. Without loosing generality, it is assumed that the given quantum circuit consists of single-qubit and two-qubit gates. If three quantum gates that act on more than two qubits, they can be decomposed into some single-qubit or two-qubit gates. Figure \ref{fig3} shows a quantum circuit with three qubits and four gates.

It is assumed a quantum circuit QC with width $W$, size $S$ and depth $D$, where:

- $W$ is the total number of qubits in the quantum circuit.

- $S$ is the total number of gates in the circuit.

- $D$ is the total time steps for executing the circuit. In each time steps, a set of gates are executed in parallel.

In the quantum circuit, qubits are shown by set $Q$ and are they numbered from one to $n$, where $i^{th}$ line from top is qubit i called $q_{i}$. The set of all gates in quantum circuit is shown by $\mathcal{G}$. Moreover, the gates are numbered in order of their executions in the quantum circuit. The gate $i^{th}$ has been shown by $g_{i}$ and means $i^{th}$ gate in scheduling algorithm for executing of gates. In scheduling algorithm, some gates are executed in parallel. For these gate, the priority of gates is arbitrary.
A distributed Quantum circuit (DQC) consists of $N$ limited capacity Quantum Circuits or partitions which are located far from each other and altogether emulate the functionality of a large quantum circuit. Partitions of DQC communicate by sending their qubits to each other using a specific quantum communication channel through teleportation \cite{Zomorodi}.

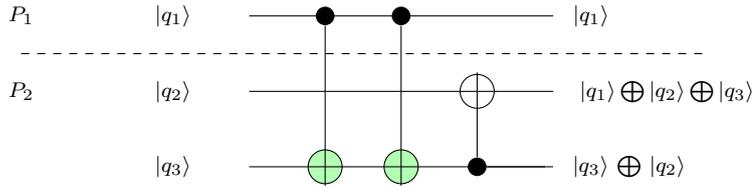
\begin{figure}[h]
\begin{center}
    \begin{tikzpicture}
    %
    % `operator' will only be used by Hadamard (H) gates here.
    % `phase' is used for controlled phase gates (dots).
    % `surround' is used for the background box.
    \tikzstyle{operatorg} = [draw,shape=circle,fill=green!30,minimum size=1.5em] 

    \tikzstyle{operator} = [draw,shape=circle,fill=white,minimum size=1.5em] 
    \tikzstyle{phase} = [draw,shape=circle,minimum size=10pt,inner sep=0pt]
    \tikzstyle{control} = [fill,shape=circle,minimum size=7pt,inner sep=0pt]
    %
    % Qubits
    \node at (0,0) (q1) {\ket{q_{1}}};
    \node at (0,-1) (q2) {\ket{q_{2}}};
    \node at (0,-2) (q3) {\ket{q_{3}}};

    \node at (-2,0) (p1) {$P_{1}$};
    \node at (-2,-1) (p1) {$P_{2}$};
    %
    % Column 2
    \node[control] (c20) at (2,0) {} ;%edge [-] (q1);
    \node[operatorg] (r22) at (2,-2) {} ;%edge [-] (q3);
    \draw[-] (2,-2.25) -- (c20);

 % Column 3
    \node[control] (c30) at (3,0) {}; %edge [-] (c20);
    \node[operatorg] (r32) at (3,-2) {}; %edge [-] (r22);
   \draw[-] (c30) -- (3,-2.25);
 % Column 4

    \node[control] (r42) at (4,-2) {} ;%edge [-] (r32);
    \node[operator] (c41) at (4,-1) {};% edge [-] (q2);
    \draw[-] (4,-0.75) -- (r42);

 \node (end1) at (5,0) {}; %edge [-] (c30);
\node (end2) at (5,-1) {} ;%edge [-] (c41);
 \node (end3) at (5,-2) {} edge [-] (r42);

 \node at (5.5,0) (q1) {\ket{q_{1}}};
    \node at (6.5,-1) (q2) {$\ket{q_{1}}\bigoplus \ket{q_{2}} \bigoplus \ket{q_{3}}$};
    \node at (6,-2) (q3) {\ket{q_{3}} $\bigoplus $ \ket{q_{2}}};
\draw[-](1,0)--(5,0);
\draw[-](1,-1)--(5,-1);
\draw[-](1,-2)--(5,-2);
\draw[dashed](-2,-0.5)--(7,-0.5);
    \end{tikzpicture}
  
  
\caption{ \small{Circuit presented in Figure\ref{fig3} is partitioned two parts:$p_{1}$ and $p_{2}$.}}
\label{fig4}
\end{center}
\end{figure}

In DQC, There are two types of quantum gates:

\begin{itemize}

\item
\textbf{Local gate:} A local gate consists of single-qubit and local CNOT gates: A single qubit has been shown by tuple $g_{i}(q_{j},p_{k})$, where $g_{i}$ is $ i^{th}$ single-qubit gate acting on $j^{th}$ qubit and $p_{k}$ is partition $k$ that gate $j^{th}$ is lactated on it. In local CNOT gates, target and control qubits are in the same partition and is shown by $g_{i}(q_{t},q_{c},p_{k})$, where $ g_{i}$ is $i^{th}$ gate, $ q_{t}$ is target qubit and $q_{c}$ is control qubit and $p_{k}$ is partition k that gate $j^{th}$ is located on it.
\item
\textbf{Global gate :} A global gate is the one whose target and control qubits are in the different partitions. This gate is shown by $g_{i}(q_{t},q_{c},p_{t},p_{c})$ where $p_{t}$ and $p_{c}$ are partitions which $q_{t}$ and $q_{c}$ are belonged on them respectively.
\end{itemize}
%Without loosing generality, it assume quantum circuit consists single-qubit or two-qubits gates. For quantum gate with more than 3 qubits, it is omitted and replaced with single-qubit or two-qubits ones.
Figure \ref{fig4} has partitioned the circuit presented in Figure \ref{fig3} to two partitions $p_{1}$ and $p_{2}$. By this partitioning, global gates are $g_{1}$ and $g_{2}$ and local gate is $g_{3}$. The set of all gates ($\mathcal{G}$)is as follows:

$\mathcal{G}=CNOT(q_{1},q_{3},p_{1},p_{2}),CNOT(q_{1},q_{3},p_{1},p_{2}),CNOT(q_{2},q_{3},p_{2})$

\section{Related work}
%Fifteen years ago, distributed quantum computing has been studied.
First ideas on distributed quantum computing were suggested by \cite{grover}, Cleve and Buhrman \cite{Cleve1997} and later by Cirac et.al. \cite{Cirac1999}. In \cite{grover} a distributed quantum system is proposed. In this system, some particles have located far from each other and sent the required data information to a base station. He divides her quantum computing into several quantum computing parts and shows how her proposed algorithm acted optimally. Grover showed by suggested distributed quantum system, computation time is faster proportional to the particle.

There are many limitations for realizing a quantum computer. Also, many numbers of qubits for building a monolithic quantum system have technological limitations. This limitation is one of reason to appear distributed quantum computing \cite{van2016path}.
Two types of communication for DQC are presented by Yepez \cite{yepez2001type}. In Type I, quantum computers use quantum communication between subsystems. In this type, each qubit may be entangled with the number of qubits. In Type II quantum computer exploits classical communication between subsystems of the distributed computer. In type II, a quantum computer consists of many quantum systems and they connect via classical channels.

An algorithm was presented by Zomorodi et al. \cite{Zomorodi} to optimize the number of qubit teleportations in a distributed quantum circuit. In this study, two spatially separated and long-distance quantum subsystem is considered. For different configurations of gate locations, the algorithm is run to calculate the minimum number of teleportations.

In \cite{Pablo}, Pablo and Chris reduced the problem to hypergraph partitioning. They represented two routines called pre- and post-processing to improve the circuit distribution. Then, they evaluated their approach on five quantum circuits and showed that the distribution cost was more than halved in comparison to the naive approach.

Another modeling of distributed quantum circuits can be found in \cite{yimsiriwattana}. In this model, non-local gates of Shor algorithm have been implemented by the distributed quantum circuit. Also, the number of teleportations is calculated but no attempt has been done to minimize the number of teleportations.

Some definition of the distributed quantum circuit has been provided by Ying and Feng \cite{ying2009algebraic}. They presented an algebraic language for modeling quantum circuits.
Van Meter et al. \cite{meter} presented a distributed quantum circuit for VBE carry-ripple adder. In this work, the VBE adder was divided into two separate quantum circuits and the circuits were communicated with each other through teleportation. So no attempt has been provided in this work to reduce the number of teleportations and there was a teleportation circuit for each global gate in the DQC. They considered two models called teledata and telegate topologies and proved that teledata is better than telegate.
Beals et al. \cite{Beals2013} presented a hypercube graph for a distributed quantum computer witch nodes connected via this graph and emulate a quantum circuit with low overhead. They showed any quantum circuit can be replaced by a DQC whose nodes are connected via a hypercube model.

Streltsov et al. \cite{streltsov} proposed a way for distributed entanglement and provided the minimum quantum cost for sending an entangled composite state in long distance. They showed the amount of entanglement sent in the total process of distribution communication may not be more than the total entanglement for sending the ancilla particle and sending back that particle.

In \cite{caleffi2018quantum} authors studied the challenges of designing quantum internet. Also they discussed that faster processing speed achieved by using connecting quantum computers via quantum internet. In another work \cite{cacciapuoti2019quantum}, the authors studied the creation of quantum internet and considered teleportation as the main protocol to transfer the information. Then they explored the challenges and open issues in the design of quantum internet.

\section{Bipartite graph partitioning }

 As stated, our new DQC model is based on graph partitioning. So, in this section, The approach of graph partitioning is discussed which has been used in this paper.
The graph partitioning problem is an interesting field which is used in the VLSI circuit design \cite{andreev}, task scheduling, clustering and social networks and many other fields \cite{bulucc2016recent}.

 Since this problem is NP-Hard \cite{andreev2006balanced}, some heuristics are used for the solution. There are many methods to solve graph partitioning such Kernighan-Lin \cite{kernighan1970efficient}, Fiduccia-Mattheyses algorithm \cite{fiduccia1982linear} as multi-level methods\cite{hendrickson1995multi,karypis1998fast,karypis1999multilevel}, spectral partitioning \cite{zare2010data,arias2011spectral} and etc.

\textit{Definition I}: Consider undirected and un weighted graph  $G= (V, E)$, where $V$ denotes the set of $n$ vertices and $E$ the set of edges. The graph partitioning problem takes a graph $G (V,E)$ as an input and a parameter $k$ giving the number of parts we wish to partition the graph into $K$ disjoint parts(sub-graph) $(V_{1},V_{2},...,V_{K})$  such that each vertex of G is contained in exactly one sub-graph and all vertices are covered. Also the Communication  cost among all of  different parts(sub-graph) is minimized, This value is calculated as follows:

\begin{equation}
\sum \limits_{i=1}^{K-1} \sum \limits_{j=i+1}^K \sum\limits_{v_{1}\in p_{i},v_{2}\in p_{j}} w({v_{1},v_{2}})
\end{equation}

Where $w(v_{1},v_{2})$ is the weight of between vertices $v_{1}$ and $v_{2}$ for all $ v_{1}\in p_{i},v_{2}\in p_{j}$. In our problem, any weigth assign to edges of graph G. Therefore, in unweigthed graphs the communication cost is the number of edges among all of  different sub-graphs $p_{i} , i=1,...K$.

Because our model is based on bipartite graphs, here some definitions related to bipartite graphs are introduced.

\textit{Definition II}: A graph $G (V, E)$ is a bigraph whose vertices can be divided into two disjoint and independent sets $X$ and $Y$ $(V=X\cup Y)$ so that each edge connects a vertex in $X$ to one in $Y$. Each set $X$ and $Y$ is called a part of the graph. This notation is presented in \cite{Thulasiraman}.

\begin {figure}[h]
\begin{center}
\tikzstyle{line} = [draw, -latex']
\tikzstyle{cir1} = [fill=blue!20,shape=circle,minimum size=15pt,inner sep=0pt] 
\tikzstyle{cir2} = [fill=yellow,shape=circle,minimum size=15pt,inner sep=0pt]

\tikzstyle{vertexs}=[draw,minimum width=0pt,minimum height=15pt]
\tikzstyle{cir3} = [fill=green!30,shape=circle,minimum size=15pt,inner sep=0pt]
\begin{tikzpicture} 

\draw [line,dashed] (0,0) ellipse (20pt and 20pt);
\draw [line,dashed] (3,0) ellipse (50pt and 10pt);
\vertexs [cir1](a1) at (0,0) []{$q_{1}$};
\vertexs [cir2](a2) at (2,0) []{$q_{2}$};
\vertexs [cir2](a3) at (4,0) []{$q_{3}$};
\vertex[cir3] (b1) at (0,2) []{$g_{1}$};
\vertex [cir3](b2) at (2,2) []{$g_{2}$};
\vertex (b3) at (4,2) []{$g_{3}$};

\path
(a1) edge (b1)
(a1) edge (b2)
(a1) edge (b2)
(a2) edge (b3)
(a2) edge (b3)
(a2) edge (b3)
(a3) edge (b1)
(a3) edge (b2)
(a3) edge (b3);
    \node at (0,-1) (p1) {$P_{1}$};
    \node at (3,-1) (p1) {$P_{2}$};
\end{tikzpicture}
\caption{\small{Bipartite graph of quantum circuit of Figure \ref{fig4}. The qubits are located in two parts:$P_{1}, P_{2}$.}}
\label{fig5}
\end{center}
\end{figure}
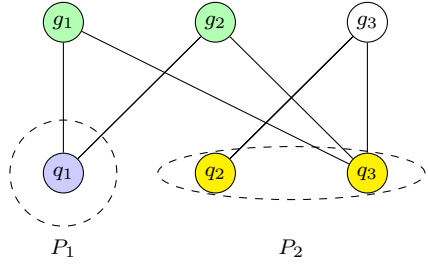

For the representation of quantum circuit by a bigraph, it is required to determine sets $X$ and $Y$ and edges between them. In our proposed model, we have considered sets $X$ and $Y$ as qubit set ($Q$) and gate set($\mathcal{G}$) respectively. The edge set of bigraph ($E$)is determined as follows:

For each $q\in Q $ and $ g\in \mathcal{G} $, there is an edge $(q,g)\in E$, if qubit $ q\in Q $ is control or target input of gate $ q \in
\mathcal{G} $ of quantum circuit. Two quantum gates describe in before, construct the edges of bigraph as follows:

\begin{itemize}
\item
\textbf{} For a single-qubit gate $g_{i}(q_{j},p_{k})$ where $ g_{i}\in Y ,q_{j} \in X$, an edge $(g_{i},q_{j})$ is added to bigraph. Also in a two-qubits gate $g_{i}(q_{t},q_{c},p_{k})$, edges$(g_{i},q_{t})$ and $(g_{i},q_{c})$ are added to bigraph.
\item
\textbf{} For a global gate $g_{i}(q_{t},q_{c},p_{t},p_{c})$ edges$(g_{i},q_{t})$ and $(g_{i},q_{c})$ are added to bigraph.
\end{itemize}
The total number of vertices in graph is $W+S$. For example, the bigraph model of quantum circuit in Figure \ref{fig4} has been shown in
Figure \ref{fig5}. In this Figure, The sets  $X$ and $Y$ of bigraph are $Q=\{q_{1},q_{2},q_{3}\}$ and $
\mathcal{G}=\{g_{1}, g_{2}, g_{3}\}$ respectively. For example $g_{1} $ has $q_{1}$ and $q_{3}$ as control and target qubits respectively. Therefore edges $(g_{1},q_{1})$ and $(g_{1},q_{3})$ are added to the bigraph. Another edges of the bigraph are added as follows:
\begin{equation}
\begin{split}
g_{1}(q_{1},q_{3},p_{1},p_{2}) \Rightarrow (g_{1},q_{1}),(g_{1},q_{3})\\
g_{2}(q_{1},q_{3},p_{1},p_{2}) \Rightarrow (g_{2},q_{1}), (g_{2},q_{3})\\
g_{3}(q_{2},q_{3},p_{2}) \Rightarrow (g_{3},q_{2}),(g_{3},q_{3}) \\
E=\{(g_{1},q_{1}),(g_{1},q_{3)},(g_{2},q_{1}), (g_{2},q_{3}),(g_{3},q_{2}),(g_{3},q_{3})\}
\end{split}
\end{equation}

As shown Figure \ref{fig4}, it is assumed that qubits are partitioned into two parts: $q_{1}$ is assigned to Part 1 and ${q_{2},q_{3}}$ are assigned to part two. As shown in Figure \ref{fig5}, the control and the target qubits of gates $g_{1}$ and $g_{2}$ are located in the different parts. Therefore, they are called global gates but gate $g_{1}$ is a local gate (having control and target qubits in the same part).

\section {Proposed Algorithm}
In this section, our proposed approach for finding the minimum number of communications in DQC is presented. It is assumed that the quantum circuit consists of single-qubit and two-qubit(CNOT) gates. The main algorithm is given in Algorithm 1 which receives the quantum circuit and number of partitions ($K$) as inputs and returns the minimum number of communication as an output.

The $main$ algorithm consists of two steps ($I$ and $II$) which are performed by $QCtoBigraph$ and $DP$ functions respectively. In Step $I$, the quantum circuit is converted to a bigraph as described in Section 5. This procedure is done by $QCtoBigraph$ function and is called in Line 4 of the $main$ algorithm.

\begin{algorithm}
{
\fontsize{9}{1}
\caption{Main algorithm}
%\label{fig4}
\begin{algorithmic}[1]

\Function {Main} {QC,K}

\State\textcolor[rgb]{0.00,0.50,1.00}{ $\triangleright$ Input: Quantum circuit (QC), The number of partitions ($K$)}

\State\textcolor[rgb]{0.00,0.50,1.00}{ $\triangleright$ Output: The minimum number of teleportations }
\State \textcolor[rgb]{0.00,0.50,1.00}{ $\triangleright$ Step I:} G=QCtoBigraph(QC);
\State \textcolor[rgb]{0.00,0.50,1.00}{ $\triangleright$ Step II:}  Number of teleportation=DP(G,K);
\EndFunction
\end{algorithmic}
}
\end{algorithm} 
\begin{algorithm}
{
\fontsize{9}{1}
\caption{This algorithm converts quantum circuit to bigraph}
%\label{fig4}
\begin{algorithmic}[2]

\Function {G=QCtoBigraph} {QC}
\State\textcolor[rgb]{0.00,0.50,1.00}{ $\triangleright$   Step I: Convert QC to Bigraph}
\State Initialize G(V,E) ,   $G.X=Q$  ,  $G.Y=\mathcal{G}$ and   $ E=\{\}$ ; \textcolor[rgb]{0.00,0.50,1.00}{ $\triangleright$ $ V=X \cup Y$}
\State\textcolor[rgb]{0.00,0.50,1.00}{ $\triangleright$  Qubits are in one part of bigraph G(part X) and gates are in other parts( part Y)  }
   \For {each $ g_{i} \in \mathcal{G}$}
         \State $c\leftarrow$ contorol qubit of $g_{i}$;
        \State $t \leftarrow$ target qubit of $g_{i}$;
         \State  Add to $E$ edges $(c,g_{i})$ and $(t,g_{i})$;
 \EndFor
\EndFunction
\end{algorithmic}
}
\end{algorithm}

$QCtoBipartite$ function is presented in Algorithm 2. This function takes the quantum circuit as an input and illustrates the bigraph as an output($G$). As stated, a bigraph has two vertex sets called $X$ and $Y$. Let $G(V, E)$ be a bigraph. In Line 3, the vertices set $X$ and $Y$ are set to $Q$ and $\mathcal{G}$ respectively and the edge $(E)$ is equal to empty. In Lines 5-8, edges are added to $E$ according to the gates of QC from left to right as mentioned in Section 5.

\begin{algorithm}
{
\fontsize{9}{1}
\caption{Dynamic programming to find the minimum number of communication}
%\label{fig4}
\begin{algorithmic}[3]

\Function{number of teleportation=DP} {G,k}    

\State\textcolor[rgb]{0.00,0.50,1.00}{ $\triangleright$ Step II: DP approach to find minimum number of  teleportation}
\State Initialize set $S$ with member of $X$ of graph $G$
\If {$k==1$ } 
\State $Return$  0;
\EndIf
\State $index$= Comupte decimal number of S; 
\State $ C[index,k]=\infty$;
\For {each $S^{'} \subset S$  }
    \State $q=connect(S^{'},S-S^{'})+DP(S-S^{'},k-1)$;
   \If  {$q\leq C[index,k]$}
       \State $C[index,k]=q$;
   \EndIf
\EndFor 
\State $Return$  $C[index,k]$;    
\EndFunction

\Function {count=Connect}{$S_{1},S_{2}$}
\State \textcolor[rgb]{0.00,0.50,1.00}{ Output:The number of global gates between $S_{1}$  and $S_{2}$}
\State $count=0$;
\For{ each $q_{i}  \in S_{1}$}
\For{  each $q_{j} \in S_{2}$}
\For { each $g_{k} \in \mathcal{G}$}
 \If{$(g_{k},q_{i}) \in E$  and $(g_{k},q_{j}) \in E$}
     \State $count++$;
\EndIf
\EndFor
\EndFor
\EndFor
\State $Return$ $count$;
\EndFunction

\end{algorithmic}
}
\end{algorithm}

In Step II, dynamic programming (DP) algorithm is presented (Algorithm 3) to find the minimum the number of communications. This function is called in Line 5 of the $Main$ algorithm.

In the first step of the DP, the optimal sub-structure must be determined and then the main optimal solution is constructed which is obtained from optimal solutions of sub-problems.

Let $T(S_{i},j)$ be the minimum number of communications for partitioning the set $S$ to $j$ parts where set $S$ consists of subset $X$ of bigraph $G$ with size $i$. In other words, subset $X$ (subset of qubit) is partitioned to $j$ parts. For the full problem, the lowest-cost way would thus be $T(S_{n},K)$.

$T(S_{i},j)$ can be defined recursively as follows:

\begin{equation}
\begin{split}
T(S_{i},j)=min_{S^{'}_{k} \subset S_{i} }( connect(S^{'}_{k}, S_{i}-S^{'}_{k})+T(S_{i}-S^{'}_{k},j-1))\\
S.t      1\leq k < i
\end{split}
\end{equation}

Let $T(S,k)$ use the function $connect(S_{1},S_{2})$. This function counts the number of global quantum gates between two-qubit sets $S_{1}$ and $S_{2}$. In other words, for each two-qubit global gate $g_{k}=(q_{t},q_{c})$, if there is $q_{t} \in S_{1}$ and $q_{c} \in S_{2}$ or conversely, this function is increased by one. Eq (9) shows this function.

\begin {equation}
 connect (S_{1},S_{2})=| Global\_gate(q_{t},q_{c})|
\end{equation}
\begin{center}
 $S.t$      ($q_{t} \in S_{1}$  and  $q_{c}\in S_{2}$)    $or$       ($q_{c} \in S_{1}$ and  $q_{t}\in S_{2}$)
\end{center}

\begin{figure}
\begin{center}
\tikzstyle{line} = [draw, -latex']

    \begin{tikzpicture}

\tikzstyle{edge_style} = [draw=black,line width=0.3,-latex']

\draw [line,dashed][rotate around={20:(-6,-3)},red] (-6,-3) ellipse (90pt and 20pt);
    %
    % Qubits
    \node at (-4,0) (q1) {$(S_{n},k)$};
 \node at (-8,-2) (q2) {$(S^{'}_{n-1},k-1)$};
 \node at (-4,-2) (q3) {$(S^{'}_{n-2},k-1)$};
 \node at (0,-2) (q4) {...};
 \node at (3,-2) (q5) {$(S^{'}_{n/2},k-1)$};
 \node at (-11,-4) (q6) {$(S^{'}_{n-2},k-2)$};
 \node at (-8,-4) (q7) {$(S^{'}_{n-3},k-2)$};
\node at (-6,-4) (q8) {...};
\node at (-4,-4) (q9) {$(S^{'}_{n-4},k-2)$};
\node at (-2,-4) (q10) {...};
\path[edge_style](q1)edge(q2);
\path[edge_style](q1)edge(q3);
\path[edge_style](q1)edge(q4);
\path[edge_style](q1)edge(q5);
\path[edge_style](q2)edge(q6);
\path[edge_style](q2)edge(q7);
\path[edge_style](q2)edge(q8);
\path[edge_style](q3)edge(q7);
\path[edge_style](q3)edge(q9);
\path[edge_style](q3)edge(q10);

    \end{tikzpicture}
  
  
\caption{\small{The recursion tree for computation of $T(S,k)$. Each node of the tree contains subset $S^{'}$ and $K$. Dashed elipse shows the subproblem overlaps.}}
\label{fig6}
\end{center}
\end{figure}
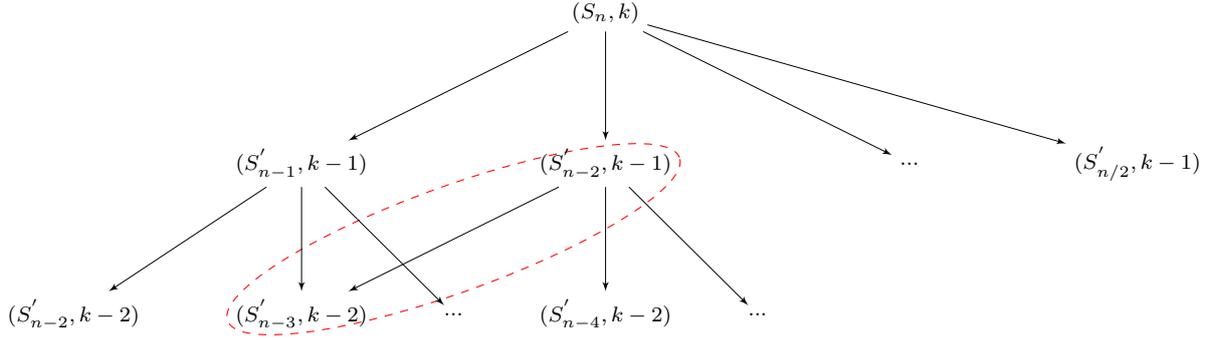

Figure \ref{fig6} shows the recursion tree of DP. In the root of tree, the main problem ($T(S_{n},K)$) is placed. This value determines the minimum number of communications for partitioning the set $S$ to $K$ parts where set $S$ consists of set $X$ of bigraph $G$ with the size $n$. In each level of tree, we compute subproblem for each subset $S^{'} \subseteq S$ and $K-1$.

Moreover, dynamic programming algorithms typically take the advantage of overlapping subproblems by solving each subproblem once and storing the solution in a table where it can be looked up when needed.
This problem has been shown in the recursion tree of Figure \ref{fig6}. It references entry $T(S^{'}_{n-3},k-2)$ many times; during computations of entries $T(S^{'}_{n-2},k-1)$ and $T(S^{'}_{n-1},k-1)$ and etc.

As a result of overlapping, we considered a table called $C$ and value of $T(S,k)$ is placed in position $C[index,k]$ so that the value of $index$ is defined as follows.

Let $b$ be a sequence of bits with size $n$. When $q_{i} \in Q$ is present in $S^{'}$, $i$th bit in $b$ becomes one; otherwise it becomes zero. Then the value of $index$ is set to the decimal value of $b$.

$DP$ function has bigraph $G$ and the number of partitions ($K$) as inputs and returns the entries of table $C$ as output by the concept of Eq (8) recursively. In the beginning of this function (Line 3), the set $S$ is initialized by the set $X$ of bigraph $G$. In Lines 4-5, if $K$ is equal to one, then the communication cost will be zero for one part. The decimal number of $S$ is computed and the value of $index$ equal to it. The minimum number of communications to partition set $S$ to $K$ parts is found among all subsets $S^{'}\subseteq S$ in Lines 8-11. Also, Function $connect(S_{1},S_{2})$ is given in Lines 13-20. This function counts the number of global gates between two sets $S_{1}$ and $S_{2}$.

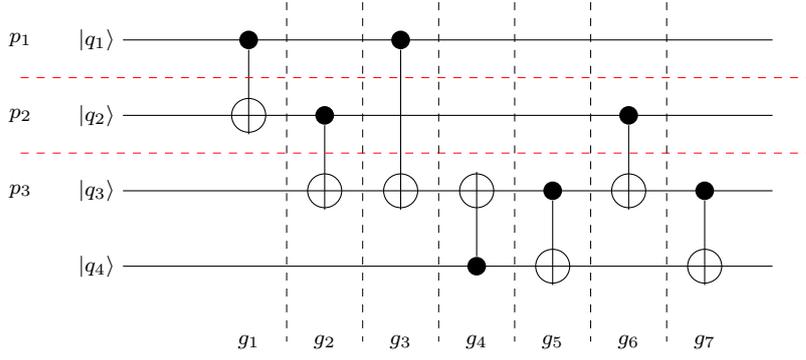
\begin{figure}[h]
\begin{center}
    \begin{tikzpicture}
    %
    % `operator' will only be used by Hadamard (H) gates here.
    % `phase' is used for controlled phase gates (dots).
    % `surround' is used for the background box.
    \tikzstyle{line1} = [draw=red,line width=0.3] 
    \tikzstyle{line2} = [draw=black,line width=0.3] 
    \tikzstyle{operator} = [draw,shape=circle,fill=white,minimum size=1.5em] 
    \tikzstyle{phase} = [draw,shape=circle,minimum size=10pt,inner sep=0pt]
    \tikzstyle{control} = [fill,shape=circle,minimum size=7pt,inner sep=0pt]
    %
    % Qubits
    \node at (0,0) (q1) {\ket{q_{1}}};
    \node at (0,-1) (q2) {\ket{q_{2}}};
    \node at (0,-2) (q3) {\ket{q_{3}}};
    \node at (0,-3) (q4) {\ket{q_{4}}};

    % Column 2
    \node[control] (c20) at (2,0) {} ;edge [-] (q1);
    \node[operator] (r21) at (2,-1) {} ;edge [-] (q2);
    \draw[-] (2,-1.25) -- (c20);

 % Column 3
    \node[control] (c31) at (3,-1) {}; %edge [-] (q2);
    \node[operator] (r33) at (3,-2) {}; %edge [-] (q3);
   \draw[-] (c31) -- (3,-2.25);

 % Column 4
    \node[control] (r40) at (4,0) {} ;%edge [-] (r32);
    \node[operator] (c42) at (4,-2) {};% edge [-] (q2);
    \draw[-] (4,-2.25) -- (r40);
    %
   % Column 5
    \node[operator] (r52) at (5,-2) {} ;%edge [-] (r32);
    \node[control] (c53) at (5,-3) {};% edge [-] (q2);
    \draw[-] (5,-1.75) -- (c53);
    %
   
% Column 6
    \node[control] (r62) at (6,-2) {} ;%edge [-] (r32);
    \node[operator] (c63) at (6,-3) {};% edge [-] (q2);
    \draw[-] (6,-3.25) -- (r62);

%Column 7
 \node[control] (r71) at (7,-1) {} ;%edge [-] (r32);
    \node[operator] (c72) at (7,-2) {};% edge [-] (q2);
    \draw[-] (7,-2.25) -- (r71);

%Column 8
 \node[control] (r82) at (8,-2) {} ;%edge [-] (r32);
    \node[operator] (c83) at (8,-3) {};% edge [-] (q2);
    \draw[-] (8,-3.25) -- (r82);

 \node (end1) at (9,0) {} edge [-] (q1);
\node (end2) at (9,-1) {} edge [-] (q2);
 \node (end3) at (9,-2) {} edge [-] (q3);
 \node (end4) at (9,-3) {} edge [-] (q4);

%part
\draw [line1,dashed](-1,-0.5)-- (9.5,-0.5);
\draw  [line1,dashed](-1,-1.5)-- (9.5,-1.5);
 \node at (-1,0)  {$p_{1}$};  \node at (-1,-1) {$p_{2}$}; \node at (-1,-2)  {$p_{3}$};

%gate
\draw [line2,dashed](2.5,0.5)-- (2.5,-4);
\draw  [line2,dashed](3.5,0.5)-- (3.5,-4);
\draw  [line2,dashed](4.5,0.5)-- (4.5,-4);
\draw  [line2,dashed](5.5,0.5)-- (5.5,-4);
\draw  [line2,dashed](6.5,0.5)-- (6.5,-4);
\draw  [line2,dashed](7.5,0.5)-- (7.5,-4);
  \node at (2,-4)  {$g_{1}$};  \node at (3,-4)  {$g_{2}$};  \node at (4,-4)  {$g_{3}$};  \node at (5,-4)  {$g_{4}$}; 
 \node at (6,-4)  {$g_{5}$};  \node at (7,-4)  {$g_{6}$};   \node at (8,-4)  {$g_{7}$}; 
    \end{tikzpicture}
\end{center}
\caption{\small{The example of quantum circuit.}}
\label{fig7}
\end{figure}
For example, we can consider bipartite graph of Figure \ref{fig7} where $G.X=\{q_{1},q_{2},q_{3},q_{4}\}$ and $ G.Y=\{g_{1},g_{2},g_{3},g_{3},g_{4},g_{5},g_{6},g_{7}\}$. It is assumed that $K=3$. The steps of the algorithm for calculating $T(\{q_{1},q_{2},q_{3},q_{4}\},3)$ are as follows:

$T(\{q_{1},q_{2},q_{3},q_{4}\},3)=$
\[
\min_{S_{'}\subseteq S}=
\begin{cases}
                                               connect(\{q_{1}\},\{q_{2},q_{3},q_{4}\})+T(\{q_{2},q_{3},q_{4}\},2)=2+2=4  \\
                                                connect(\{q_{2}\},\{q_{1},q_{3},q_{4}\})+T(\{q_{1},q_{3},q_{4}\},2)=3+1=4\\
                                                connect(\{q_{3}\},\{q_{1},q_{2},q_{4}\})+T(\{q_{1},q_{2},q_{4}\},2)=6+0=6\\
                                                connect(\{q_{4}\},\{q_{1},q_{2},q_{3}\})+T(\{q_{1},q_{2},q_{3}\},2)=3+2=5\\
                                                connect(\{q_{1},q_{2}\},\{q_{3},q_{4}\})+T(\{q_{3},q_{4}\},2)=3+3=6\\
                                                connect(\{q_{1},q_{3}\},\{q_{2},q_{4}\})+T(\{q_{2},q_{4}\},2)=6+0=6\\
                                               connect(\{q_{1},q_{4}\},\{q_{2},q_{3}\})+T(\{q_{2},q_{3}\},2)=5+2=7\\
 \end{cases}
\]

As shown above, for obtaining the final solution of $T(\{q_{1},q_{2},q_{3},q_{4}\},3)$, it is required to solve $T(\{q_{2},q_{3},q_{4}\},2)$, $T(\{q_{1},q_{3},q_{4}\},2),...$ recursively. Table \ref{table1} indicates these results computed by $DP$ function for this circuit. In this table, rows indicate the number of partitions and the columns represent the set of qubits participate in partitioning. Also, the decimal value of each qubit set is given in the first column.

In this example, the minimum number of communications, which is four, occurs for $\{\{q_{1}\},\{q_{2}\},\{q_{3},q_{4}\}\}$. The solution shows that $\{q_{1}\}$ is placed in partition 1 and $\{q_{2},q_{3},q_{4}\}$  are partitioned to two parts recursively. By solving $T(\{q_{2},q_{3},q_{4}\},2)$, qubit sets $\{q_{2}\}$ and $\{q_{3},q_{4}\}$  are assigned to parts 2 and 3 recursively.

Let us consider the steps of the gate executions according to this partitioning. The algorithm starts with the first gate in $
\mathcal{G}$ i.e. $g_{1}(q_{1},q_{2},p_{1},p_{2})$ which is a global gate. For executing this gate, qubit $q_{1}$ in $p_{1}$ is teleported to $p_{2}$, and the number of communication is increased by one, and then step by step all other gates are executed and removed from the list. Other steps of running gates are as follows:	

\begin{itemize}
\item
\textbf{}$g_{2}(q_{2},q_{3} ,p_{2},p_{3})$ is a global gate and $q_{2}$ is teleported to part three and executed there.
\item
\textbf{}$g_{3}(q_{1},q_{3},p_{1},p_{3})$ is a global gate because its target input is in part three and its control input is in part one.  Therefore, $q_{1}$ in part one is teleported to part three for executing $g_{3}$.
		\item
\textbf{}$g_{4}$ and $ g_{5}$  are local gates and are executed in part three.
\textbf{}$g_{6}$ and $g_{7}$ are global and local gates respectively and are executed the same as other gates.
\end{itemize}

\begin{table}[H]
\caption{\small{Table $C$ obtained from the $DP$ function for the circuit of Figure \ref{fig7}.}}
\label{table1}
\footnotesize{
\begin{tabular}{|c|p{0.3cm}|p{0.3cm}|p{0.4cm}|p{0.3cm}|c|c|c|c|c|c|c|c|c|c|c|}
\hline
k&1&2&3&4&5&6&7&8&9&10&11&12&13&14&15\\
&\{1\}&$\{2\}$&\{2,1\}&\{3\}&\{3,1\}&\{3,2\}&\{3,2,1\}&\{4\}&\{4,1\}&\{4,2\}&\{4,2,1\}&\{4,3\}&\{4,3,1\}&\{4,3,2\}&\{4,3,2,1\}\\
\hline

3&N.A&N.A&N.A&N.A&N.A&N.A&N.A&N.A&N.A&N.A&N.A&N.A&N.A&N.A&2\\
\hline
2&N.A&N.A&1&N.A&1&2&2&N.A&0&0&0&3&1&2&2\\
\hline

1&0&0&0&0&0&0&0&0&0&0&0&0&0&0&0\\
\hline
\end{tabular}
}
\end{table}

\section{Experimental Results}
We implemented our algorithm in MATLAB on a workstation with 4GB RAM and 0.5 GHz CPU to find the best partitioning with an optimized number of teleportation. Many different quantum circuits were used for comparing the performance of our algorithm with other approaches. These quantum circuits are as follows:
\paragraph{-}
Quantum fourier transform (QFT) \cite{Nielsen10}: in quantum computing, the quantum fourier transform (QFT) is a linear transformation on quantum bits. QFT is used in some quantum algorithms such as Shor's algorithm. The quantum gates used in the implementation of this algorithm are the Hadamard gate and the controlled phase gate $R_{m}$ as described in Section 3.
\paragraph{-}
Binary welded tree (BWT) \cite{Childs}: it consists of two balanced binary trees of the height $n$ with the $2^n$ leaves of the left tree identified with $2^n$ leaves of the right tree. In this circuit, Toffoli gates are replaced with CNOT gates.

\paragraph{-}
Ground state estimation (GSE) \cite{whit2011}: twice the default number of basic functions and occupied orbitals.

\paragraph{-}
Another set of test samples for quantum circuits was taken from Revlib \cite{Wille} library which is an online resource of benchmarks. We used some of them such as: Alu\_primitive, Parity, Flip\_flop, Sym9\_147.

\paragraph{-}
To compare the results with the work in \cite{Zomorodi}, we used the same quantum circuit example of \cite{Zomorodi}.

For comparision with Pablo method, we used the $ratio (R)$ used in \cite{Pablo} as follows:
\begin{equation}
R=\frac{Number\_teleportations}{2*Number\_qubits} 
\end{equation}

As stated before, each teleportation comprises two qubits and each qubit is located in different part. Therefore half of teleportation is related to the number of qubits. For this purpose, number two is used in the fraction of this equation. Having $R > 1$ means that the number of teleportations is greater than the number of qubits and some qubits are teleported more than once. Therefore this distribuition has not act well comared to $R<1$.

\begin{figure}[h]
  \centering
  \begin{subfigure}[b]{0.5\linewidth}
    \includegraphics[width=\linewidth]{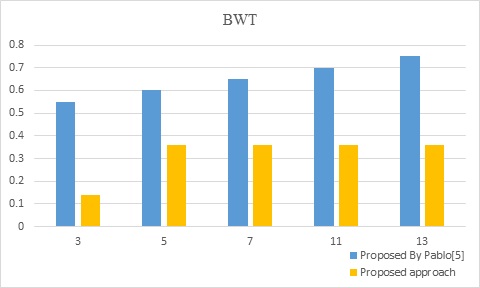}
    \caption{BWT sample}
  \end{subfigure}
  \begin{subfigure}[b]{0.5\linewidth}
    \includegraphics[width=\linewidth]{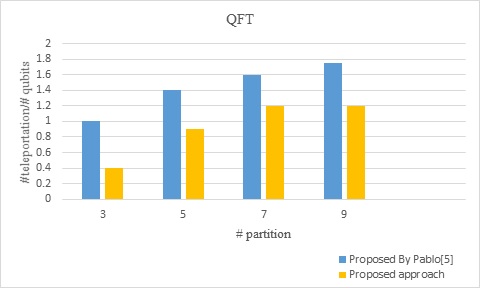}
    \caption{QFT sample}
  \end{subfigure}
\begin{subfigure}[b]{0.5\linewidth}
    \includegraphics[width=\linewidth]{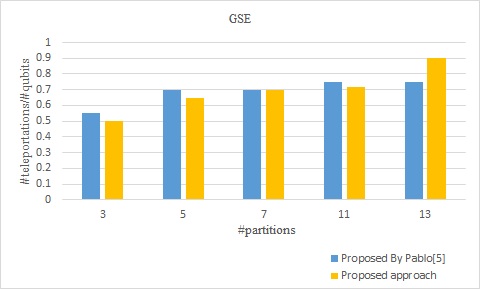}
    \caption{GSE sample}
  \end{subfigure}
  \caption{\small{Each bar shows the ratio $(R)$ between the number of teleportations and number of qubits. $X$ and $Y$ axes show the  number of partitions and $R$ respectively.}}
  \label{fig8}
\end{figure}

Figure \ref{fig8} shows this ratio $(R)$ for various number of partitions $(K)$ in comparison with Pablo \cite{Pablo} for three quantum circuits: BWT (Figure \ref{fig8}.a), QFT (Figure \ref{fig8}.b), and GSE (Figure \ref{fig8}.c) circuits. In comparison with \cite{Pablo}, the parameter $R$ is better except for GSE circuit which did not distribute well for $K=13$. Also proposed method produced the same $R$ for $K=7$ in GSE circuit. By comparing the values in Table \ref{table2}, QFT for $K=3$ did not produce good result by the method presented in \cite{Pablo} and required several qubits for communication. Also, for $K>=5$, QFT required more qubits than the number of communications $(R > 1)$. But in our proposed approach, for $K=3, 5$ the distribution is performed better: The ratio was less than one (R $< 1$). The exact values of $R$ are given in Table \ref{table2}.

Table \ref{table3} shows the minimum number of communications for parity\_247, Sym9\_147, Flip\_flop, Alu\_primitive, Alu\_primitive\_opt, and random circuit example of \cite{Zomorodi}. In this table, the number of qubits, gates, and partitions are given for each sample. We compared the results of the proposed approach with two proposed methods presented in \cite{Zomorodi} and \cite{ZomorodiE} in terms of the teleportation cost (TC).

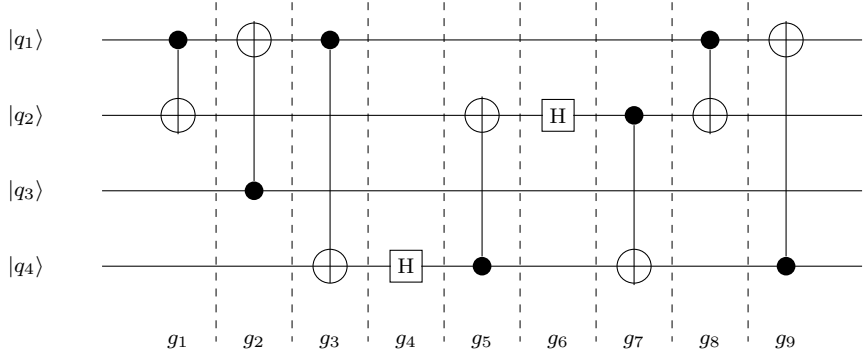
\begin{figure}
\begin{center}
     \begin{tikzpicture}
    %
    % `operator' will only be used by Hadamard (H) gates here.
    % `phase' is used for controlled phase gates (dots).
    % `surround' is used for the background box.
    \tikzstyle{operator} = [draw,shape=circle,fill=white,minimum size=1.5em] 
    \tikzstyle{phase} = [draw,shape=circle,minimum size=10pt,inner sep=0pt]
    \tikzstyle{control} = [fill,shape=circle,minimum size=7pt,inner sep=0pt]
   \tikzstyle{hadamard} = [draw,shape=rectangle,minimum size=12pt,inner sep=0pt]
       \tikzstyle{line2} = [draw=black,line width=0.3] 
    %
    % Qubits
    \node at (-1,0) (q1) {\ket{q_{1}}};
    \node at (-1,-1) (q2) {\ket{q_{2}}};
    \node at (-1,-2) (q3) {\ket{q_{3}}};
    \node at (-1,-3) (q4) {\ket{q_{4}}};

    %
    % Column 1
    \node[control] (c10) at (1,0) {} ;%edge [-] (q1);
    \node[operator] (r12) at (1,-1) {} ;%edge [-] (q3);
    \draw[-] (1,-1.25) -- (c10);

 % Column 2
    \node[control] (c22) at (2,-2) {}; %edge [-] (c20);
    \node[operator] (r20) at (2,0) {}; %edge [-] (r22);
   \draw[-] (c22) -- (2,0.25);
 % Column 3

    \node[control] (r30) at (3,0) {} ;%edge [-] (r32);
    \node[operator] (c33) at (3,-3) {};% edge [-] (q2);
    \draw[-] (3,-3.225) -- (r30);
    %
   
   %column 4
   
   \node [hadamard] (h4)  at (4,-3){H};
   %column 5
   
       \node[control] (r53) at (5,-3) {} ;%edge [-] (r32);
    \node[operator] (c51) at (5,-1) {};% edge [-] (q2);
    \draw[-] (5,-0.75) -- (r53);
    
    % column 6;
    
   \node [hadamard] (h6)  at (6,-1){H};
    
    %column 7
    
       \node[control] (c71) at (7,-1) {} ;%edge [-] (r32);
    \node[operator] (r73) at (7,-3) {};% edge [-] (q2);
    \draw[-] (7,-3.25) -- (c71);
    % column 8

       \node[control] (c80) at (8,0) {} ;%edge [-] (r32);
    \node[operator] (r81) at (8,-1) {};% edge [-] (q2);
    \draw[-] (8,-1.25) -- (c80);
    
    %column 9
     
       \node[control] (c93) at (9,-3) {} ;%edge [-] (r32);
    \node[operator] (r90) at (9,0) {};% edge [-] (q2);
    \draw[-] (9,0.25) -- (c93);

\draw[-](0,0)--(10,0);
\draw[-](0,-1)--(5.8,-1);\draw[-](6.2,-1)--(10,-1);
\draw[-](0,-2)--(10,-2);
\draw[-](0,-3)--(3.8,-3);\draw[-](4.2,-3)--(10,-3);

\draw [line2,dashed](1.5,0.5)-- (1.5,-4);
\draw [line2,dashed](2.5,0.5)-- (2.5,-4);
\draw  [line2,dashed](3.5,0.5)-- (3.5,-4);
\draw  [line2,dashed](4.5,0.5)-- (4.5,-4);
\draw  [line2,dashed](5.5,0.5)-- (5.5,-4);
\draw  [line2,dashed](6.5,0.5)-- (6.5,-4);
\draw  [line2,dashed](7.5,0.5)-- (7.5,-4);
\draw [line2,dashed](8.5,0.5)-- (8.5,-4);

  \node at (1,-4)  {$g_{1}$};  \node at (2,-4)  {$g_{2}$};  \node at (3,-4)  {$g_{3}$};  \node at (4,-4)  {$g_{4}$}; 
 \node at (5,-4)  {$g_{5}$};  \node at (6,-4)  {$g_{6}$};   \node at (7,-4)  {$g_{7}$}; 
 \node at (8,-4)  {$g_{8}$};   \node at (9,-4)  {$g_{9}$}; 
    \end{tikzpicture}
\caption{Sampel quantum circuit. The figure was reproduced from \cite{Zomorodi}.}
\label{fig9}
\end{center}
\end{figure}

Let us consider the sample circuit of \cite{Zomorodi} in figure \ref{fig9}. This sample circuit has been reproduced from \cite{Zomorodi}, (their figure 4) for reference. In \cite{Zomorodi}, minimum number of communications, which is four, occurs for Config-Arr=\{11000\} where the first and second global gates are executed in $P_{1}$ and other global gates are executed in partition $P_{0}$. In our model, $\{q_{1},q_{2},q_{3}\}$ and $\{q_{3}\}$ are located in $P_{0}$ and $P_{1}$ respectively for $K=2$. The minimum number of communications was two and the steps of running gates are shown in Table \ref{table4}.
The model of \cite{Zomorodi} had some limitations: in the beginning of their algorithm, partitions were fixed and they did not afford to find optimized partitions. Therefore, their space model was limited to two pre-defined partitions and they considered different configurations for this pre-defined partitioning.

\begin{table}[H]
\begin{center}
\caption{\small{The ratio between number of teleportaions (halves) and number of qubits for $K=3,5,..,13$ for QFT, BWT and GSE circuits in comparison with \cite{Pablo}(RP)}}
\label{table2}
\footnotesize{
\begin{tabular}{cp{0.3cm}p{0.3cm}p{0.01cm}p{0.3cm}p{0.3cm}p{0.01cm}p{0.3cm}p{0.3cm}p{0.01cm}p{0.3cm}p{0.3cm}p{0.01cm}p{0.3cm}p{0.3cm}p{0.01cm}p{0.3cm}p{0.3cm}}
%&\multicolumn{2}{r}{ k=3}&\multicolumn{2}{r}{ 5} &\multicolumn{2}{r}{ 7}& \multicolumn{2}{r}{ 9}&\multicolumn{2}{r}{ 11} &\multicolumn{2}{r}{ 13} \\
&K=3&&&K=5&&&K=7&&&K=9&&&K=11&&&K=13&\\
\cline{2-3} \cline{5-6}\cline{8-9} \cline{11-12}\cline{14-15}\cline{17-18}
Circuit&R&RP&&R&RP&&R&RP&&R&RP&&R&RP&&R&RP\\
\hline
BWT&0.12&0.55&&0.35&0.6&&0.35&0.64&&0.7&0.65&&0.35&0.7&&0.36&0.77\\

QFT&0.4&1&&0.85&1.4&&1.2&1.58&&1.2&1.75&&1.6&1.75&&1.75&1.8\\

GSE&0.5&0.54&&0.64&0.7&&0.7&0.7&&0.79&0.8&&0.72&0.74&&0.9&0.74\\

\end{tabular}
}

\end{center}
\end{table}

\begin{table}[H]
\begin{center}
\caption{\small{Comparison of the proposed approach (P) with \cite{Zomorodi} and \cite{ZomorodiE}.}}
\label{table3}
\footnotesize{
\begin{tabular}{c|cccccc}
Circuit&\#of qubits &\# of gates&$K$&TC [42]&TC [1]&TC (P)\\
\hline

pariaty\_247&17&16&2&2&2&2\\
Sym9\_147&12&108&2&48& N.A.&8\\
Flip\_flop&8&30&3&N.A.&N.A.&8\\
Alu\_primitive&6&21&2&20&18&6\\
Alu\_primitive\_opt&6&21&2&10&10&6\\
Figure 4 of \cite{Zomorodi}&4&7&2&4&4&2\\
\end{tabular}
} 
\end{center}
\end{table}

\begin{table}[H]
\begin{center}
\caption{\small{The steps of the proposed algorithm for the random circuit of \cite{Zomorodi}. $L$ and $G$ stand for local and global gates respectively}.}

\begin{tabular}{c|cc}

\# of Gate& Gate\_name& Type of gate\\
\hline
$g_{1}$&$CNOT(q_{1},q_{2},p_{0})$& $L$\\

$g_{2}$&$CNOT(q_{3},p_{1},q_{1},p_{0})$&$G$\\

$g_{3}$&$CNOT(q_{1},q_{4},p_{0}$&$L$\\

$g_{4}$&$H(q_{4},p_{0})$&$L$\\

$g_{5}$&$CNOT(q_{2},q_{4},p_{0})$&$L$\\

$g_{6}$&$H(q_{2},p_{0})$&$L$\\

$g_{7}$&$CNOT(q_{2},q_{4},p_{0})$&$L$\\

$g_{8}$&$CNOT(q_{1},q_{2})$&$L$\\

$g_{9}$&$CNOT(q_{1},q_{4},p_{0})$&$L$\\

\end{tabular}

\label{table4}
\end{center}
\end{table}

\section{Conclusion}
Teleportation is a costly operation in quantum computation and it is very important to minimize the number of this operation in computations. In this study, an algorithm was proposed for distributing quantum circuits to optimize the number of teleportations between qubits. The proposed algorithm consisted of two steps: in the first step, the quantum circuit was converted to a bipartite graph (bigraph) and in the next step by a dynamic programming approach, bigraph was partitioned into $K$ parts. Finally, compared with previous works in \cite{Zomorodi},\cite{Pablo}, and \cite{ZomorodiE}, it was shown that the proposed approach yielded the better or the same results for benchmark circuits.

%\begin{acknowledgements}
%If you'd like to thank anyone, place your comments here
%and remove the percent signs.
%\end{acknowledgements}

% Authors must disclose all relationships or interests that 
% could have direct or potential influence or impart bias on 
% the work: 
%
% \section*{Conflict of interest}
%
% The authors declare that they have no conflict of interest.

% BibTeX users please use one of
%\bibliographystyle{spbasic}      % basic style, author-year citations
%\bibliographystyle{spmpsci}      % mathematics and physical sciences
%\bibliographystyle{spphys}       % APS-like style for physics
%\bibliography{}   % name your BibTeX data base

% Non-BibTeX users please use

\bibliographystyle{spphys}
\bibliography{ref}

\begin{thebibliography}{10}
\providecommand{\url}[1]{{#1}}
\providecommand{\urlprefix}{URL }
\expandafter\ifx\csname urlstyle\endcsname\relax
  \providecommand{\doi}[1]{DOI \discretionary{}{}{}#1}\else
  \providecommand{\doi}{DOI \discretionary{}{}{}\begingroup
  \urlstyle{rm}\Url}\fi

\bibitem{Zomorodi}
M.~Zomorodi-Moghadam, M.~Houshmand, M.~Houshmandi, Theoretical Physics
  \textbf{57}(3), 848–861 (2018)

\bibitem{van2010}
R.~Van~Meter, T.D. Ladd, A.G. Fowler, Y.~Yamamoto, International Journal of
  Quantum Information \textbf{8}, 295 (2010)

\bibitem{krojanski2004scaling}
H.G. Krojanski, D.~Suter, Physical review letters \textbf{93}(9), 090501 (2004)

\bibitem{nickerson2013topological}
N.H. Nickerson, Y.~Li, S.C. Benjamin, Nature communications \textbf{4}, 1756
  (2013)

\bibitem{Pablo}
P.~Andres-Martinez, Theoretical computer science \textbf{410}(26), 2489 (2018)

\bibitem{Meter06}
R.V. Meter, M.~Oskin, ACM Journal on Emerging Technologies in Computing Systems
  (JETC \textbf{2}, 2006 (2006)

\bibitem{meter}
V.~Meter, W.~Munro, K.~Nemoto, K.M. Itoh, ACM Journal on Emerging Technologies
  in Computing Systems (JETC) \textbf{3}, 2 (2008)

\bibitem{Bennett}
C.H. Bennett, G.~Brassard, C.~Crepeau, R.~Jozsa, A.~Peres, W.K. Wootters,
  Physical Review Letters \textbf{70}, 1895 (1993)

\bibitem{Whit}
M.~Whitney, N.~Isailovic, Y.~Patel, J.~Kubiatowicz, Physical Review Letters
  \textbf{85}(26), 1330 (2000)

\bibitem{Nielsen10}
M.A. Nielsen, I.L. Chuang, \emph{Quantum computation and quantum information},
  10th edn. (Cambridge University Press, 2010)

\bibitem{wootteres}
W.K. Wootters, W.H. Zurek, nature \textbf{299}, 802 (1982)

\bibitem{farhi2000quantum}
E.~Farhi, J.~Goldstone, S.~Gutmann, M.~Sipser, arXiv preprint quant-ph/0001106
  (2000)

\bibitem{zomorodi2014synthesis}
M.~Zomorodi-Moghadam, M.A. Taherkhani, K.~Navi, in \emph{Transactions on
  Computational Science XXIV} (Springer, 2014), pp. 74--91

\bibitem{deutsch}
D.E. Deutsch, Proceedings of the Royal Society of London. A. Mathematical and
  Physical Sciences \textbf{425}(1868), 73 (1989)

\bibitem{weinstein2014steane}
Y.~Weinstein, S.~Buchbinder, Journal of Modern Optics \textbf{61}(1), 49 (2014)

\bibitem{zomorodi2016rotation}
M.~Zomorodi-Moghadam, K.~Navi, Journal of Circuits, Systems and Computers
  \textbf{25}(12) (2016)

\bibitem{grover}
L.K. Grover, arXiv preprint quant-ph/9704012  (1997)

\bibitem{Cleve1997}
R.~Cleve, H.~Buhrman, Physical Review A \textbf{56}, 1201 (1997)

\bibitem{Cirac1999}
J.~Cirac, A.~Ekert, S.~Huelga, C.~Macchiavello, Physical Review A \textbf{59},
  4249 (1999)

\bibitem{van2016path}
R.~Van~Meter, S.J. Devitt, Computer \textbf{49}(9), 31 (2016)

\bibitem{yepez2001type}
Yepez, Jeffrey, International Journal of Modern Physics C \textbf{12}(09), 1273
  (2001)

\bibitem{yimsiriwattana}
Yimsiriwattana, Anocha, L.~Jr, S.~J, arXiv preprint quant-ph/0403146  (2004)

\bibitem{ying2009algebraic}
M.~Ying, Y.~Feng, IEEE Transactions on Computers \textbf{58}(6), 728 (2009)

\bibitem{Beals2013}
R.~Beals, S.~Brierley, O.~Gray, A.W. Harrow, S.~Kutin, N.~Linden, D.~Shepherd,
  M.~Stather, Proceedings of the Royal Society A: Mathematical, Physical and
  Engineering Sciences \textbf{469}(2153), 20120686 (2013)

\bibitem{streltsov}
A.~Streltsov, H.~Kampermann, D.~Bru{\ss}, Physical review letters \textbf{108},
  250501 (2012)

\bibitem{caleffi2018quantum}
M.~Caleffi, A.S. Cacciapuoti, G.~Bianchi, in \emph{Proceedings of the 5th ACM
  International Conference on Nanoscale Computing and Communication} (2018),
  pp. 1--4

\bibitem{cacciapuoti2019quantum}
A.S. Cacciapuoti, M.~Caleffi, F.~Tafuri, F.S. Cataliotti, S.~Gherardini,
  G.~Bianchi, IEEE Network  (2019)

\bibitem{andreev}
K.~Andreev, H.~Racke, Theory of Computing Systems \textbf{39}(6), 929 (2006)

\bibitem{bulucc2016recent}
A.~Bulu{\c{c}}, H.~Meyerhenke, I.~Safro, P.~Sanders, C.~Schulz, in
  \emph{Algorithm Engineering} (Springer, 2016), pp. 117--158

\bibitem{andreev2006balanced}
K.~Andreev, H.~Racke, Theory of Computing Systems \textbf{39}(6), 929 (2006)

\bibitem{kernighan1970efficient}
B.W. Kernighan, S.~Lin, Bell system technical journal \textbf{49}(2), 291
  (1970)

\bibitem{fiduccia1982linear}
C.M. Fiduccia, R.M. Mattheyses, in \emph{19th Design Automation Conference}
  (IEEE, 1982), pp. 175--181

\bibitem{hendrickson1995multi}
B.~Hendrickson, R.W. Leland,

\bibitem{karypis1998fast}
G.~Karypis, V.~Kumar, SIAM Journal on scientific Computing \textbf{20}(1), 359
  (1998)

\bibitem{karypis1999multilevel}
G.~Karypis, R.~Aggarwal, V.~Kumar, S.~Shekhar, IEEE Transactions on Very Large
  Scale Integration (VLSI) Systems \textbf{7}(1), 69 (1999)

\bibitem{zare2010data}
H.~Zare, P.~Shooshtari, A.~Gupta, R.R. Brinkman, BMC bioinformatics
  \textbf{11}(1), 403 (2010)

\bibitem{arias2011spectral}
E.~Arias-Castro, G.~Chen, G.~Lerman, et~al., Electronic Journal of Statistics
  \textbf{5}, 1537 (2011)

\bibitem{Thulasiraman}
M.N.S.S. CK.~Thulasiraman, \emph{Theory and Algorithms} (1994)

\bibitem{Childs}
A.M. Childs, R.~Cleve, E.~Deotto, E.~Farhi, S.~Gutmann, D.A. Spielman, STOC '03
  Proceedings of the thirty-fifth annual ACM symposium on Theory of computing
  \textbf{410}(26), 59 (2003)

\bibitem{whit2011}
J.D. Whitfield, J.~Biamonte, A.~Aspuru-Guzik, Molecular Physics
  \textbf{109}(5), 735 (2011)

\bibitem{Wille}
R.~Wille, D.~Grobe, L.~Teuber, G.~Dueck, R.~Drechsler, Multiple-Valued Logic,
  IEEE International Symposium on \textbf{0}, 220 (2008)

\bibitem{ZomorodiE}
M.Z.M. Zahra~Mohammadi, M.~Houshmand, M.~Houshmandi, Arxiv  (2019)

\end{thebibliography}

\end{document}